\documentclass[twocolumn,prl,showpacs]{revtex4-1}
\usepackage{graphicx}
\usepackage{amsmath}
\usepackage{amsfonts}
\usepackage{mathtools}
\usepackage{subfigure}

\begin{document}
\title{Stochastic discrete time crystals: Entropy production and subharmonic synchronization}

\author{Lukas Oberreiter$^1$, Udo Seifert$^1$, and Andre C. Barato$^2$}
\affiliation{$^1$ II. Institut f\"ur Theoretische Physik, Universit\"at Stuttgart, 70550 Stuttgart, Germany\\
$^2$ Department of Physics, University of Houston, Houston, Texas 77204, USA}

\parskip 1mm
\def\d{{\rm d}}
\def\Ps{{P_{\scriptscriptstyle \hspace{-0.3mm} s}}}
\def\MF{{\mbox{\tiny \rm \hspace{-0.3mm} MF}}}
\def\ts{\tau_{\textrm{sig}}}
\def\tos{\tau_{\textrm{osc}}}
\begin{abstract}
Discrete time crystals are periodically driven systems that display spontaneous symmetry 
breaking of time translation invariance in the form of indefinite subharmonic oscillations. We introduce 
a thermodynamically consistent  model for a discrete time crystal and analyze it using the framework 
of stochastic thermodynamics. In particular, we evaluate the rate of energy dissipation
of this many body system of interacting noisy subharmonic oscillators in contact with a heat bath. Our model 
displays the novel phenomena of subharmonic synchronization, which corresponds to collective 
subharmonic oscillations of the individual units.  
\end{abstract}
\pacs{05.70.Ln, 02.50.Ey}
% Explanation of PACS numbers:
% 05.70.Ln: Nonequilibrium and irreversible thermodynamics
% 02.50.Ey: Stochastic processes 

\maketitle
%==========================================================================
%\section{Introduction}
%==========================================================================

Time crystals \cite{sach17,khem19} are a phase of matter first proposed by Wilczek \cite{wilc12}. They are closed  equilibrium systems with a 
time-independent Hamiltonian that displays oscillations in time, which  
corresponds to a spontaneous breaking of time translation symmetry. The name is chosen in analogy to  
crystals, which display spontaneous symmetry breaking of spatial translation symmetry due to the emergence of a
periodic arrangement of atoms in space. Shortly after this proposal, general proofs 
that time crystals could not be realized in closed many-body quantum systems with short-range 
interactions  were provided \cite{brun13,wata15}. However, quite recently, it has been demonstrated that a time crystal 
in a system with long range interactions is possible \cite{kozi19}. 

Compared to this initial proposal by Wilczek, a different kind of breaking of time translation symmetry
 happens in discrete time crystals (DTCs) \cite{sach15,khem16,else16,pizz19}. 
These are nonequilibrium quantum systems with a time-periodic Hamiltonian, for which breaking of time translation 
symmetry is manifested in the occurrence of subharmonic oscillations with a period longer than the period of the Hamiltonian. 
These discrete time crystals are not in contact with a heat bath, hence, they do not 
dissipate energy. They typically rely on disorder and localization to avoid a stationary state of infinite 
temperature, which would not support time crystalline order \cite{khem16,else16,moes17}. Interestingly, DTCs have 
been realized in experiments \cite{zhan17,choi17}. 

For such DTCs, coupling the system to an external reservoir can destroy the DTC phase \cite{laza17}. 
Nevertheless, open systems in contact with an external reservoir allow for a broader range of mechanisms, which 
do not rely on disorder and localization, that do lead to a DTC phase \cite{yao18,gong18,wang18,gamb19,gamb19b,heug19}. 
In fact, the onset of subharmonic oscillations in dynamical systems under periodic driving has been known for quite some time \cite{gold18}. 
However, the amount of energy dissipated by a DTC as an open system has not been evaluated yet. 

In this Letter, we introduce a thermodynamically consistent model for a classical stochastic many-body DTC in contact with 
a heat bath. Our model falls within the theoretical framework of stochastic thermodynamics \cite{seif12,jarz11,broe15}. 
As one consequence, we can evaluate the rate of entropy production, which quantifies how much energy the system dissipates. 
We show that the average of this quantity and its fluctuations can be used to identify the transition to a DTC phase.
 
The mechanism that leads to subharmonic oscillations in our model is different from DTCs in open systems that have been proposed 
so far. We consider a many-body system for which each isolated unit displays a finite number of coherent subharmonic 
oscillations that fades away after some time due to noise \cite{ober19}. By introducing interactions between these units, we show that 
for an interaction strength above a certain critical value the number of coherent subharmonic oscillations 
diverges  in the thermodynamic limit, which is a signature of a DTC phase. 

The mean field version of our model displays a new phenomenon that we call 
subharmonic synchronization. Standard synchronization \cite{gupt18} is a fundamental phenomena in physics, 
whereby coupled oscillators display collective oscillations. For the subharmonic 
synchronization observed here, periodically driven oscillators display collective 
subharmonic oscillations. 

A surprising result is obtained with the 2D version of the model. It
does not display subharmonic synchronization, similar to related models for standard  
synchronization that do not display synchronization in 2D \cite{wood06,wood06a}. However, we show that 
the 2D model still exhibits a DTC phase, characterized by indefinite subharmonic oscillations
in the thermodynamic limit.

%==========================================================================
%\section{Model definitions}
%==========================================================================

\begin{figure}
\includegraphics[width=75mm]{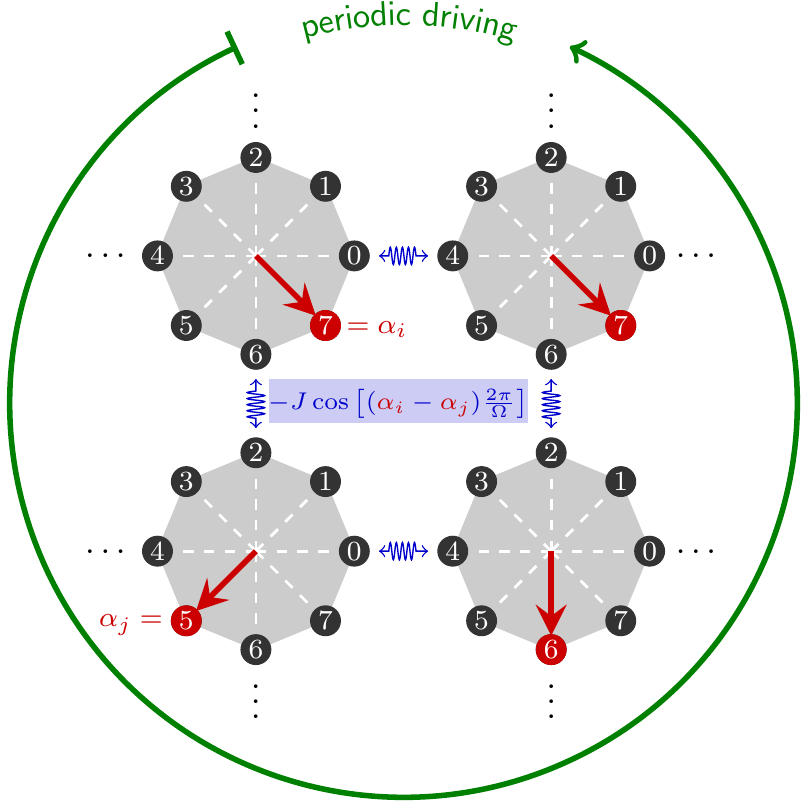}
\vspace{-2mm}
\caption{(Color online) Sketch of the 2D model.}
\label{fig1} 
\end{figure}

Each unit of our model, which is illustrated in Fig. \ref{fig1}, is a clock with $\Omega\ge 3$ states labeled by $\alpha=0,1,\ldots,\Omega-1$.
For a single unit, the transition rate 
from state $\alpha$ to $\alpha+1$ is  
\begin{equation}
w_\alpha^+(t)= k\textrm{e}^{E_\alpha(t)-B_\alpha(t)}, 
\end{equation}
while the transition rate from state $\alpha$ to $\alpha-1$  is 
\begin{equation}
w_{\alpha}^-(t)= k\textrm{e}^{E_{\alpha}(t)-B_{\alpha-1}(t)},
\end{equation}
where the parameter $k$ sets the time-scale. 
The time-periodic energy of state $\alpha$ is $E_\alpha(t)$, 
the time-periodic energy barrier between states $\alpha$ and $(\alpha+1)\mod \Omega$
is $B_\alpha(t)$. Boltzmann's constant $k_B$ and temperature $T$ are set to $k_B= T=1$ throughout. 
For $t\in [0,\tau]$, where $\tau$ is the period, the energy and energy barriers are given by 
\begin{equation}
E_\alpha(t)= [\ln(c)/\Omega][(\alpha+\lfloor{\Omega t/\tau \rfloor}-1) \mod \Omega],
\label{Ealpha}
\end{equation}
and
\begin{equation}
B_\alpha(t)= [\ln(c)/\Omega][\Omega-1+(\alpha+\lfloor{\Omega t/\tau \rfloor})\mod \Omega], 
\label{Balpha}
\end{equation}
where $c$ is a positive constant.

The model for a single unit has been analyzed in \cite{ober19}. In a certain deterministic limit, for which
$k\gg \tau^{-1}$ and $c\to \infty$, this clock  displays indefinite subharmonic oscillations with 
period $(\Omega-1)\tau$. Let us explain the case $\Omega=3$ in this limit. The period has three parts. During
the first part of the period only the transition rate from state $\alpha=0$ to state $\alpha=1$ is finite (and much 
larger than $\tau^{-1}$), with the other transition rates arbitrarily small. During the second part of the period only the transition rate 
from state  $\alpha=2$ to state $\alpha=0$ is finite. During the third part of the period only the transition rate 
from state  $\alpha=1$ to state $\alpha=2$ is finite. If we consider an initial condition $\alpha=0$,
at the end of the first period the state of the system will be $\alpha=2$. The system then starts the second period 
at $\alpha=2$ and finishes this period at state $\alpha=0$. Consequently, the system oscillates subharmonically between 
states $\alpha=0$ and $\alpha=2$ in the subsequent periods, with state $\alpha=1$ absent in the oscillations. 
The period of this subharmonic oscillation is $(\Omega-1)\tau=2\tau$. A similar argument holds for arbitrary $\Omega$, 
with the state $\alpha=1$ absent in the subharmonic oscillations.

For a single unit, these subharmonic oscillations are indefinite only in this peculiar deterministic limit. 
For finite values of $k$ and $c$, thermal fluctuations destroy the coherence of the subharmonic
oscillations, i.e., two point correlation functions in time display subharmonic oscillations 
that decay exponentially \cite{ober19}.  

The many-body system consists of $N$ such units. The state of unit $i=1,2,\ldots,N$ is denoted by  $\alpha_i$ and the 
state of the system by $\vec{\alpha}=(\alpha_1,\alpha_2,\ldots,\alpha_N)$. These units 
interact with a time-independent interaction energy 
\begin{equation}
V_{\vec{\alpha}}= -\mathcal{J}\sum_{i=1}^N\sum_{j} \cos[2\pi(\alpha_i-\alpha_j)/\Omega]/2. 
\label{eqinteraction2D}
\end{equation}
For the mean field variant, the sum in $j$ is over all units from $j=1$ to $j=N$ and $\mathcal{J}=J/N$. For the 2D variant, the sum in $j$ is over four nearest neighbors, $\mathcal{J}=J$, 
and we consider periodic boundary conditions. The time-periodic Hamiltonian of the model is 
$\sum_{i=1}^NE_{\alpha_i}(t)+V_{\vec{\alpha}}$, where $E_{\alpha_i}(t)$ is given by Eq. \eqref{Ealpha}.
The transition rate from a state $\vec{\alpha}=(\ldots,\alpha_i,\ldots)$ to state $\vec{\alpha}'=(\ldots,\alpha_i+1,\ldots)$
is given by 
\begin{equation}
w_{\vec{\alpha}_i}^+(t)= k\textrm{e}^{\theta_{\vec{\alpha},\vec{\alpha}'}(V_{\vec{\alpha}}-V_{\vec{\alpha}'})+E_{\alpha_i}(t)-B_{\alpha_i}(t)}, 
\label{eqfw}
\end{equation}
where $\theta_{\vec{\alpha},\vec{\alpha}'}= 1$ if $V_{\vec{\alpha}'}\ge V_{\vec{\alpha}}$ and $\theta_{\vec{\alpha},\vec{\alpha}'}= 0$ if $V_{\vec{\alpha}'}< V_{\vec{\alpha}}$.
The transition rate from state $\vec{\alpha}=(\ldots,\alpha_i,\ldots)$ to state $\vec{\alpha}'=(\ldots,\alpha_i-1,\ldots)$
is given by 
\begin{equation}
w_{\vec{\alpha}_i}^-(t)= k\textrm{e}^{\theta_{\vec{\alpha},\vec{\alpha}'}(V_{\vec{\alpha}}-V_{\vec{\alpha}'})+E_{\alpha_i}(t)-B_{\alpha_i-1}(t)}.
\label{eqbc}
\end{equation}

We have performed continuous-time Monte Carlo simulations of this model using the Gillespie algorithm \cite{gill77}. 
The parameters of the model are set to $k=40$, $c=10^4$ and the period is $\tau=1$. The number of states of each unit is 
$\Omega=8$. The basic phenomena we investigate is whether, for an interaction strength $J$ above a certain critical value, subharmonic oscillations become coherent in 
the thermodynamic limit, which is a signature of the onset of a DTC phase. Changing the parameters that have been fixed leads 
to results that are quantitatively different but have the same physical features discussed below.

The following observables characterize this system. First, the order parameter for the synchronization of the different clocks reads \cite{gupt18} 
\begin{equation}
r(t)\equiv N^{-1}\left|\sum_{i=1}^N \textrm{e}^{2\pi \mathrm{i} \alpha_i(t)/\Omega}\right|,
\label{eqrk}
\end{equation}
where $\alpha_i(t)$ is the state of unit $i$ at time $t$. Since we are interested in subharmonic oscillations we consider the stroboscopic 
time $n$, with $r_n\equiv r(n\tau)$. The quantity $r_n$ as a function of $n$ reaches a stationary value denoted by $r$. If the clocks do not synchronize then $r$ approaches 
$(\Omega-1)^{-1}$ rather than going to $0$. This nonzero $r$ is due to the following fact. For the case of no interactions ($J=0$),
the clocks oscillate subharmonically in a completely desynchronized fashion. However, as explained above, state $\alpha=1$ is not part of these 
oscillations, which leads to $r\to (\Omega-1)^{-1}$ for this case. Therefore, if the subharmonic clocks synchronize,
then $r> (\Omega-1)^{-1}$.  

%The average of 
%$r$ over stochastic trajectories is denoted by $\langle r\rangle$. If the clocks do not synchronize then $\langle r\rangle$ approaches 
%$(\Omega-1)^{-1}$ rather than going to $0$. This nonzero $r$ is due to the following fact. For the case of no interactions ($J=0$),
%the clocks oscillate subharmonically in a completely desynchronized fashion. However, as explained above, state $\alpha=1$ is not part of these 
%oscillations, which leads to $\langle r\rangle\to (\Omega-1)^{-1}$ for this case. Therefore, if the subharmonic clocks synchronize,
%then $\langle r\rangle> (\Omega-1)^{-1}$.  

Second, the number of coherent oscillations is quantified by the correlation function $C(t)$, which is 
the density of clocks in state $\alpha_i=0$ at time $t$ given that at time $0$ all clocks $i=1,2,\ldots,N$ are in state $\alpha_i=0$. The stroboscopic quantity $C_n=C(n\tau)$ has 
oscillations that decay exponentially in $n$. The period of oscillations and the decay time are written $n_{\textrm{osc}}$ and $n_{\textrm{dec}}$, respectively. 
The number of coherent subharmonic oscillations is defined as 
\begin{equation}
\mathcal{R}\equiv 2\pi n_{\textrm{dec}}/n_{\textrm{osc}}.
\label{eqrc}
\end{equation}   
The factor $2\pi$ in this definition is related to the fact that $\mathcal{R}$ can be defined 
as the ratio of the imaginary and real parts of an eigenvalue of the fundamental matrix \cite{ober19}.

Third, the amount of energy dissipation is quantified by the thermodynamic rate of entropy production \cite{seif12}. 
This observable quantifies the energetic cost of the DTC; it is the rate of work exerted on the system due the periodic driving. 
For a stochastic trajectory with total time $T$ exhibiting $M$ jumps, $\vec{\alpha}^{(1)}\to\vec{\alpha}^{(2)}\to\ldots\to\vec{\alpha}^{(M)}$,
the entropy change is  
\begin{equation}
X_\sigma\equiv \sum_{m=1}^{M-1}\ln(w_{\vec{\alpha}^{(m)}\to \vec{\alpha}^{(m+1)}}/w_{\vec{\alpha}^{(m+1)}\to \vec{\alpha}^{(m)}}),
\label{eqrc2}
\end{equation}   
where $w_{\vec{\alpha}^{(m)}\to \vec{\alpha}^{(m+1)}}$ is the transition rate from state $\vec{\alpha}^{(m)}$ to state $\vec{\alpha}^{(m+1)}$. 
The average rate of entropy production is $\sigma\equiv \langle X_\sigma\rangle/T$, where the brackets denote an average over stochastic 
trajectories. Fluctuations of the entropy production are quantified by  the Fano factor $F_\sigma\equiv \left(\langle X_\sigma^2\rangle-\langle X_\sigma\rangle^2\right)/\langle X_\sigma\rangle$.
Both quantities, $\sigma$ and $F_\sigma$, are formally defined in the limit $T\to\infty$.

%==========================================================================
%\section{MF results }
%==========================================================================

\begin{figure*}[t]
\subfigure[]{\includegraphics[width=59mm]{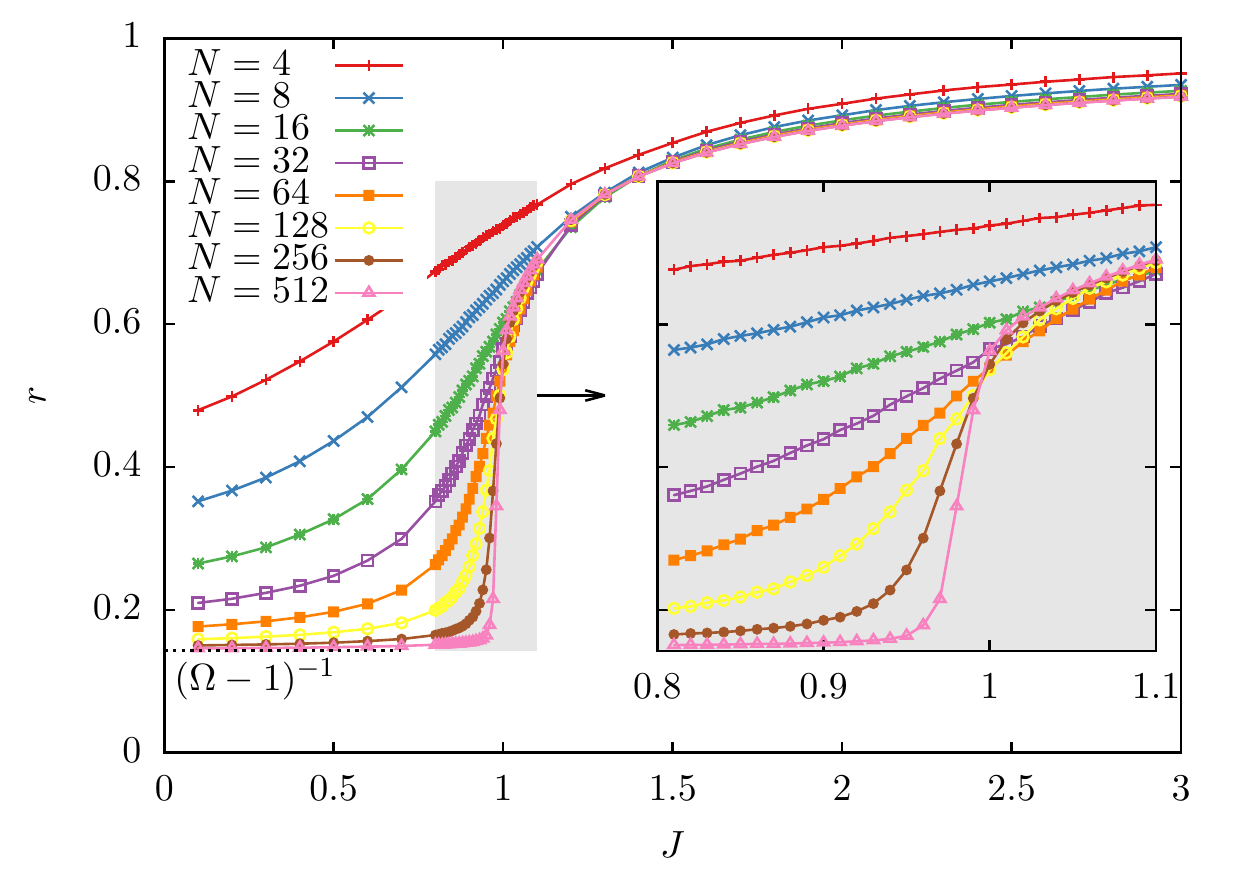}\label{fig2a}}
\subfigure[]{\includegraphics[width=59mm]{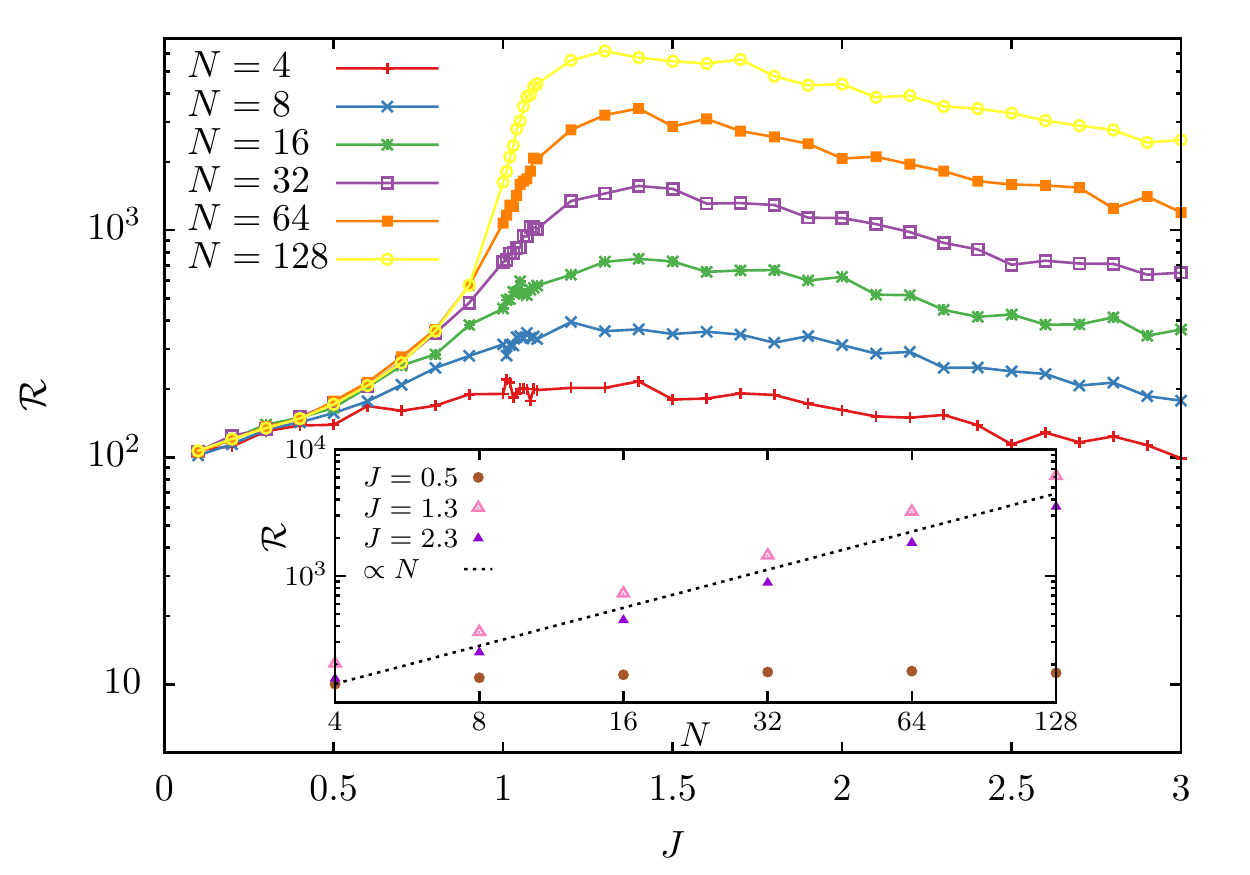}\label{fig2b}}
\subfigure[]{\includegraphics[width=59mm]{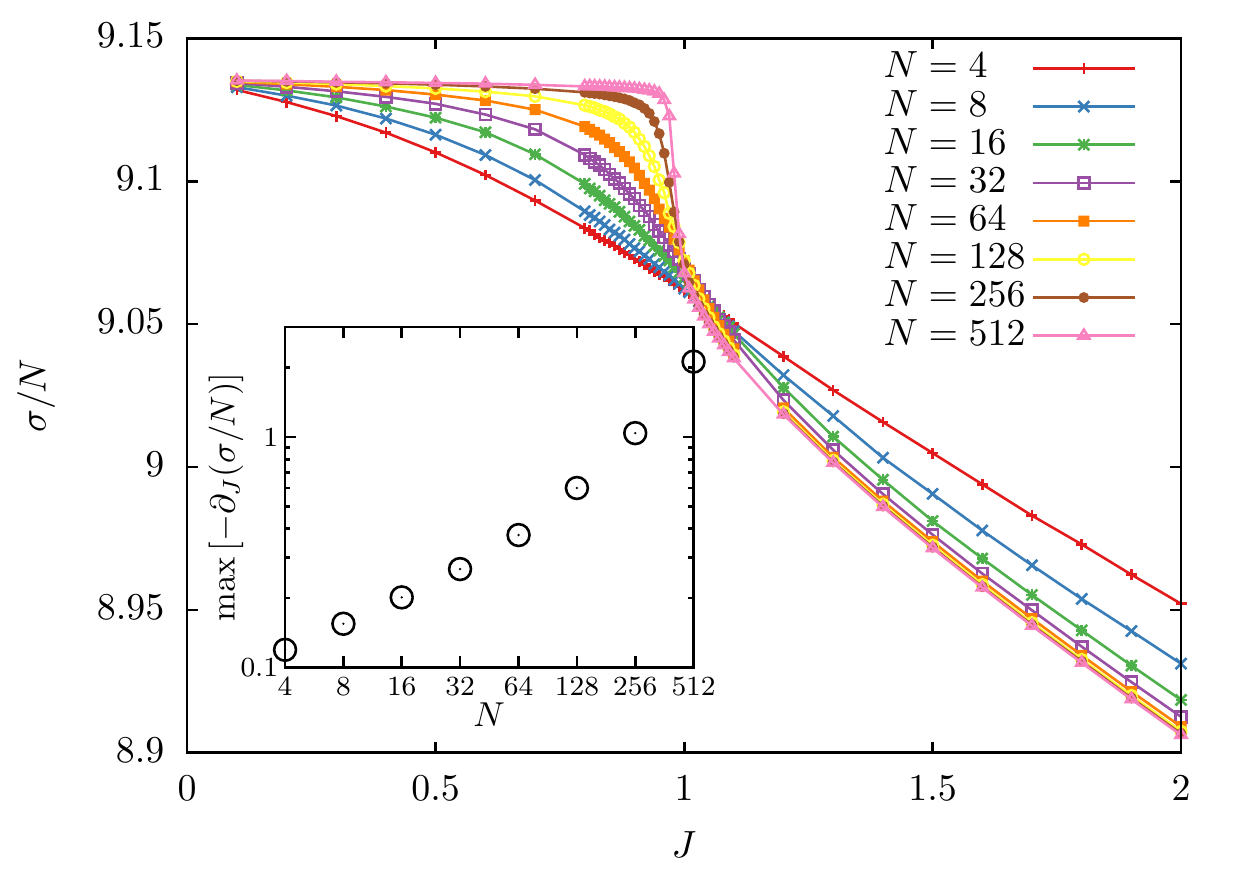}\label{fig2c}}
\vspace{-2mm}
\caption{(Color online) Observables for the mean field variant as functions of the interaction strength $J$. (a) Order parameter $r$. (b) Number of coherent oscillations $\mathcal{R}$. 
(c) Rate of entropy production per unit $\sigma/N$.}   
\label{fig2} 
\end{figure*}

We first analyze the mean field model. The results for the order parameter $r$ are shown in Fig. \ref{fig2a}. 
They indicate that $r>(\Omega-1)^{-1}$ for $J>J_c\simeq 0.975$ in the thermodynamic limit. This mean field model displays the novel phenomena of subharmonic synchronization. It differs
from related models for standard synchronization without periodic drive \cite{wood06,wood06a,herp18}. In these models each unit is a biased random walk on a circle with $\Omega$ states. 
The bias is generated by a fixed thermodynamic force such as the free energy of adenosine thriphosphate hydrolysis. In contrast, in our model each unit is a 
periodically driven clock that displays noisy subharmonic oscillations.

The scaling of the number of coherent oscillations $\mathcal{R}$ with the system size $N$ is shown in Fig. \ref{fig2b}. Below the critical point $\mathcal{R}$ 
saturates. Hence, subharmonic oscillations do not last indefinitely in the thermodynamic limit but fade away after some transient. 
Above the critical point, $\mathcal{R}$ diverges with system size as a power law, with an exponent compatible with $1$. For $J\ge J_c$,
subharmonic oscillations thus become indefinite in the limit $N\to \infty$, which corresponds to a DTC phase. 

Since our model for a DTC is thermodynamically consistent, we can evaluate how much energy this DTC dissipates using Eq. \eqref{eqrc2}. 
In Fig. \ref{fig2c}, we show the rate of entropy production per unit $\sigma/N$ as a function of the interaction strength $J$ for 
different values of $N$. The maximum of the first derivative of $\sigma/N$ with respect to $J$ as a function of $N$ seems to follow a power law, 
which indicates that this derivative diverges in the limit $N\to \infty$. With the values of $N$ that were accessible with our simulations we 
were not able to determine the exponent reliably. Our result indicates that in the thermodynamic limit there is a discontinuity in the rate of entropy
production per unit, $\sigma/N$. 

%\begin{figure}
%\includegraphics[width=75mm]{./sigma_J_inset_MF.pdf}
%\vspace{-2mm}
%\caption{Rate of entropy production per unit $\sigma/N$ as a function of the the  interaction strength $J$ for the mean field model.
%}
%\label{fig3} 
%\end{figure}

%==========================================================================
%\section{2D Results }
%==========================================================================

\begin{figure*}[t]
\subfigure[]{\includegraphics[width=59mm]{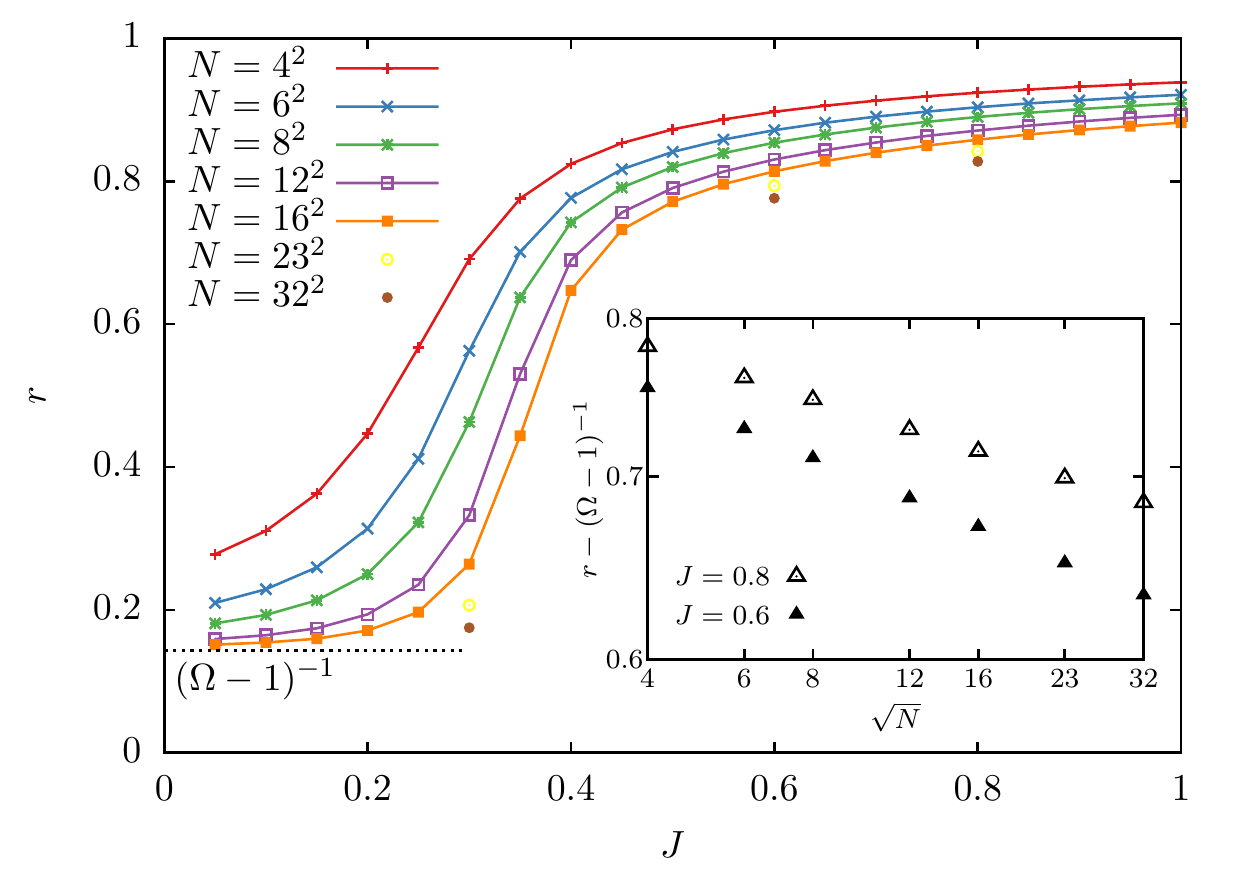}\label{fig4a}}
\subfigure[]{\includegraphics[width=59mm]{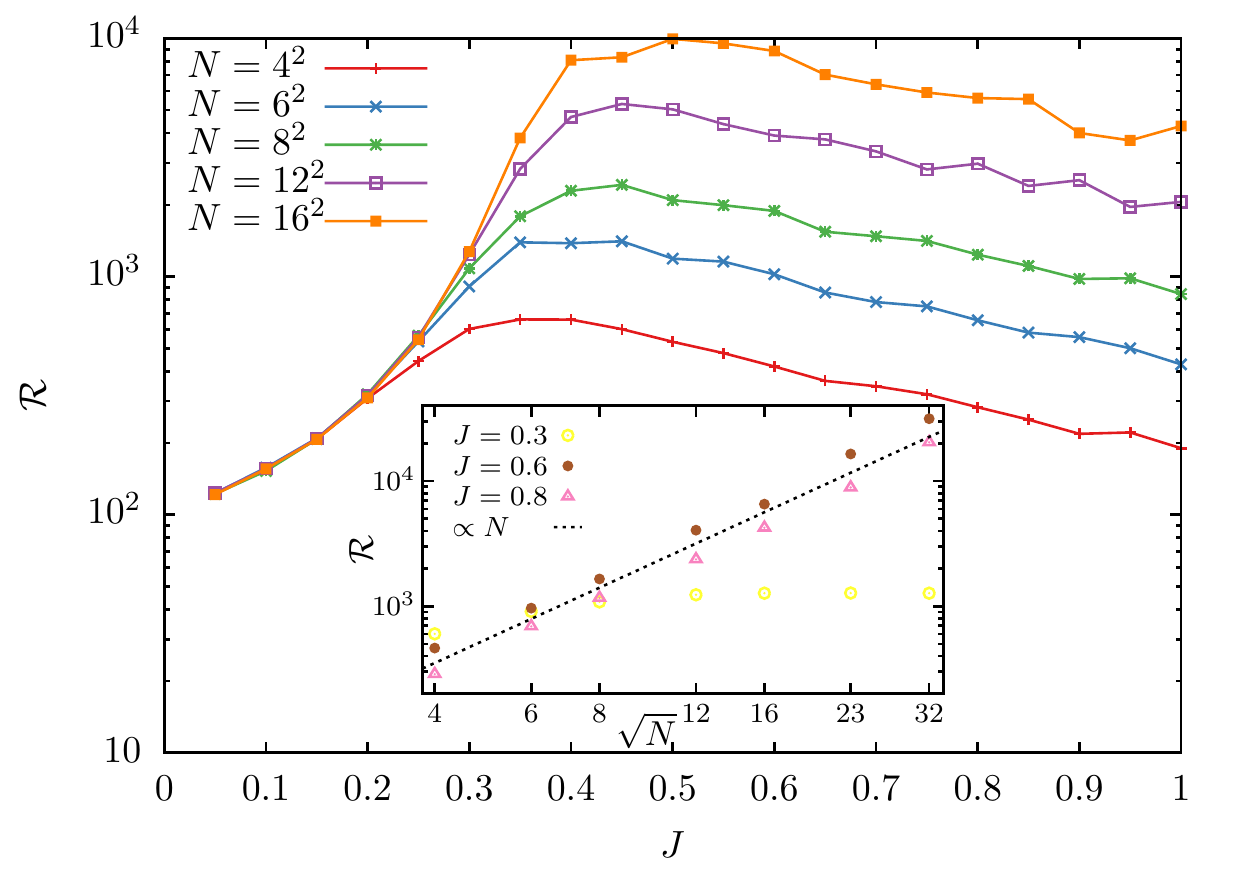}\label{fig4b}}
\subfigure[]{\includegraphics[width=59mm]{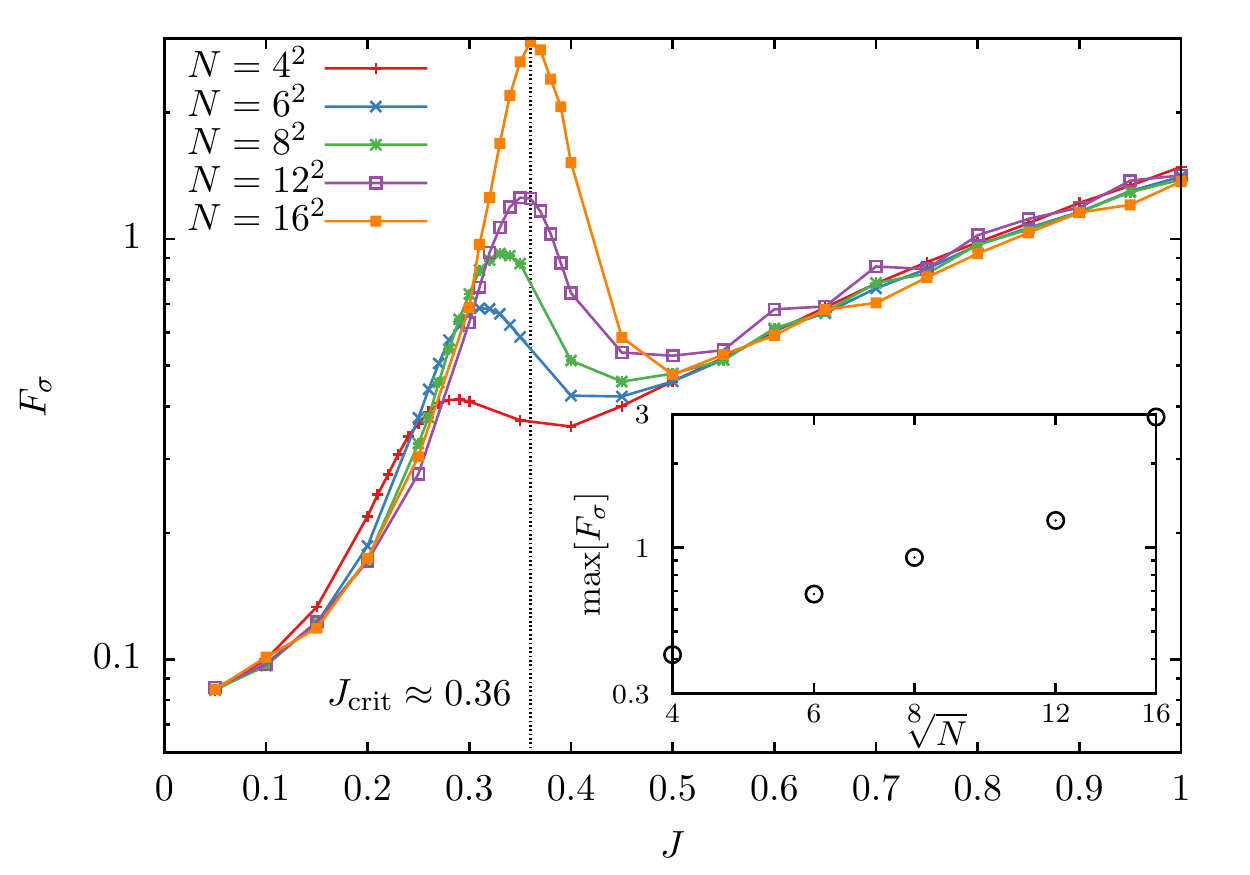}\label{fig4c}}
\vspace{-2mm}
\caption{(Color online) Observables for the 2D variant as functions of the interaction strength $J$. (a) Order parameter $r$. (b) Number of coherent oscillations $\mathcal{R}$. (c) Fano factor $F_\sigma$.}   
\label{fig3} 
\end{figure*}

For the 2D model we obtain results that are  qualitatively different from the results for the mean field model. As shown 
in Fig. \ref{fig4a}, even for regions where the order parameter $r$
seems to be larger than $(\Omega-1)^{-1}$ in a finite system, $r-(\Omega-1)^{-1}$ decays to zero as power law 
with system size. Hence, there is no subharmonic synchronization in the 2D model, with $r\to (\Omega-1)^{-1}$
for any value of $J$ in the thermodynamic limit.  

Interestingly, even though the 2D model does not display subharmonic synchronization,  it 
does still display a DTC phase. For $J$ above a certain critical value $J_c\simeq 0.36$, the number of coherent 
oscillations diverges with system size as a power law, as shown in Fig. \ref{fig4b}. Similar to the 
mean field version, the exponent is approximately $1$, independent of the value of $J$. 

%Another signature of this phase transition to a DTC phase is the divergence of the fluctuations of the order parameter, as quantified by 
%$\xi\equiv N\langle r^2\rangle-\langle r\rangle^2$, in the limit $N\to \infty$. As shown in Fig. \ref{fig5}, the maximum of 
%this quantity diverges with $N$. We were not able to determine the exponent reliably with the system sizes accessible with our simulations.

For the rate of entropy production in the 2D model, we cannot identify any non-analytical behavior of $\sigma/N$ or its first derivative at criticality within our numerics.
Non-analytical behavior of higher order derivatives cannot be ruled out. However, we can observe the signature of a phase transition in the fluctuations of the entropy production, 
as quantified by the Fano factor $F_\sigma$. As shown in Fig. \ref{fig4c}, the maximum of $F_\sigma$ diverges with system size, which indicates that the Fano factor 
diverges at criticality in the thermodynamic limit. We could not determine the exponent associated with  this divergence reliably. Divergence of this Fano factor at criticality has also 
been observed in models for biochemical oscillations \cite{nguy18}.      

\begin{figure}
\includegraphics[width=75mm]{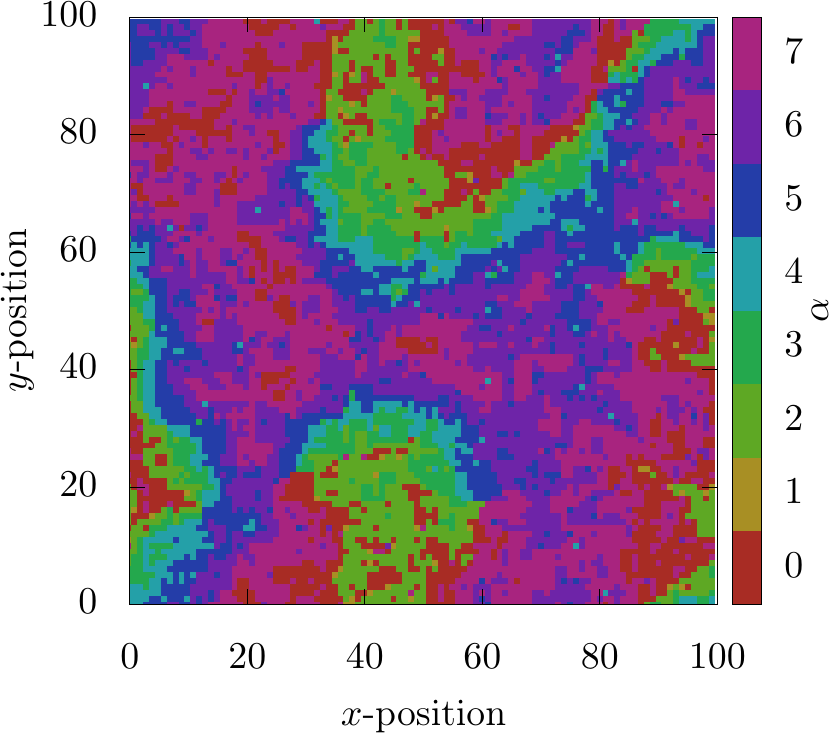}
\vspace{-2mm}
\caption{(Color online) Long-range order in the 2D model. The axes represent the 
spatial coordinates of the 2D lattice. The colors represent the state of an unit $\alpha_i=0,1,\ldots,7$.
The parameters are $N=100^2$ and $J=0.8$. This picture was taken for the stroboscopic time $n=1000$.}
\label{fig4} 
\end{figure}

Further evidence of a phase transition in the 2D model is shown in Fig. \ref{fig4}, which is a snapshot of the state 
of the system. This picture shows that there is long-range order above the critical point with the 
formation of islands with a certain orientation. This result is similar to the observation of a "Kosterlitz-Thousless-type" phenomena 
in a 2D model of interacting noisy oscillators (without periodic driving) \cite{wood06,wood06a}.

%\begin{figure}
%\includegraphics[width=77mm]{./chi_J_2D.pdf}
%\vspace{-2mm}
%\caption{Fluctuations of the order parameter $\langle r^2\rangle-\langle r\rangle^2$  as a function of the interactions strength $J$. The 
%inset shows the power law behavior of the maximum of $\langle r^2\rangle-\langle r\rangle^2$ as a function of $N$.
%}
%\label{fig5} 
%\end{figure}

%\begin{figure}
%\includegraphics[width=77mm]{./2D_sigma_J.pdf}
%\vspace{-2mm}
%\caption{Rate of entropy production per unit $\sigma/N$ as a function of the the  interaction strength $J$ for the 2D model.
%(We need the right figure here).}
%\label{fig6} 
%\end{figure}

%==========================================================================
%\section{Conclusion }
%==========================================================================

In summary, we have introduced a paradigmatic model for a stochastic many-body system in contact with a 
heat bath that displays a DTC phase. The novel phenomena of subharmonic 
synchronization, whereby periodically driven oscillators display synchronized subharmonic 
oscillations, was found with the mean-field version. For the 2D variant, there 
is no synchronization, however, there is a rich phenomenology 
with a phase transition to a DTC phase, which is characterized by the divergence of the 
number of coherent subharmonic oscillations and by spatial long-range order. We expect that future work 
on the 2D model will lead to a systematic understanding of this phenomenology.

We have calculated the rate of entropy production of a DTC with our thermodynamically 
consistent model. For the mean field model, the rate of entropy production has a 
discontinuity at the critical point, dropping to a lower 
value when the system crosses to the DTC phase. For the 2D model, while we could not 
observe any non-analytical behavior of the rate of entropy production, the Fano factor 
that quantifies fluctuations of the entropy production diverges at the critical point.

Time crystals are related to the idea of perpetual motion \cite{khem19}. We have considered 
an open system with a well defined second law of thermodynamics, which rules out 
perpetual motion. As one intriguing application, our results open up the possibility of building a 
model for a subharmonic heat engine that breaks time translation symmetry. It 
would be interesting to investigate whether power and efficiency of such a heat 
engine is bounded by the relations that have been found for cyclic 
and steady state heat engines \cite{espo09,shir16,piet18,koyu19}.

\bibliographystyle{apsrev4-1}

\bibliography{refs} 

\begin{thebibliography}{34}%
\makeatletter
\providecommand \@ifxundefined [1]{%
 \@ifx{#1\undefined}
}%
\providecommand \@ifnum [1]{%
 \ifnum #1\expandafter \@firstoftwo
 \else \expandafter \@secondoftwo
 \fi
}%
\providecommand \@ifx [1]{%
 \ifx #1\expandafter \@firstoftwo
 \else \expandafter \@secondoftwo
 \fi
}%
\providecommand \natexlab [1]{#1}%
\providecommand \enquote  [1]{``#1''}%
\providecommand \bibnamefont  [1]{#1}%
\providecommand \bibfnamefont [1]{#1}%
\providecommand \citenamefont [1]{#1}%
\providecommand \href@noop [0]{\@secondoftwo}%
\providecommand \href [0]{\begingroup \@sanitize@url \@href}%
\providecommand \@href[1]{\@@startlink{#1}\@@href}%
\providecommand \@@href[1]{\endgroup#1\@@endlink}%
\providecommand \@sanitize@url [0]{\catcode `\\12\catcode `\$12\catcode
  `\&12\catcode `\#12\catcode `\^12\catcode `\_12\catcode `\%12\relax}%
\providecommand \@@startlink[1]{}%
\providecommand \@@endlink[0]{}%
\providecommand \url  [0]{\begingroup\@sanitize@url \@url }%
\providecommand \@url [1]{\endgroup\@href {#1}{\urlprefix }}%
\providecommand \urlprefix  [0]{URL }%
\providecommand \Eprint [0]{\href }%
\providecommand \doibase [0]{http://dx.doi.org/}%
\providecommand \selectlanguage [0]{\@gobble}%
\providecommand \bibinfo  [0]{\@secondoftwo}%
\providecommand \bibfield  [0]{\@secondoftwo}%
\providecommand \translation [1]{[#1]}%
\providecommand \BibitemOpen [0]{}%
\providecommand \bibitemStop [0]{}%
\providecommand \bibitemNoStop [0]{.\EOS\space}%
\providecommand \EOS [0]{\spacefactor3000\relax}%
\providecommand \BibitemShut  [1]{\csname bibitem#1\endcsname}%
\let\auto@bib@innerbib\@empty
%</preamble>
\bibitem [{\citenamefont {Sacha}\ and\ \citenamefont
  {Zakrzewski}(2017)}]{sach17}%
  \BibitemOpen
  \bibfield  {author} {\bibinfo {author} {\bibfnamefont {K.}~\bibnamefont
  {Sacha}}\ and\ \bibinfo {author} {\bibfnamefont {J.}~\bibnamefont
  {Zakrzewski}},\ }\href {\doibase 10.1088/1361-6633/aa8b38} {\bibfield
  {journal} {\bibinfo  {journal} {Rep. Prog. Phys.}\ }\textbf {\bibinfo
  {volume} {81}},\ \bibinfo {pages} {016401} (\bibinfo {year}
  {2017})}\BibitemShut {NoStop}%
\bibitem [{\citenamefont {Khemani}\ \emph {et~al.}(2019)\citenamefont
  {Khemani}, \citenamefont {Moessner},\ and\ \citenamefont {Sondhi}}]{khem19}%
  \BibitemOpen
  \bibfield  {author} {\bibinfo {author} {\bibfnamefont {V.}~\bibnamefont
  {Khemani}}, \bibinfo {author} {\bibfnamefont {R.}~\bibnamefont {Moessner}}, \
  and\ \bibinfo {author} {\bibfnamefont {S.~L.}\ \bibnamefont {Sondhi}},\
  }\href@noop {} {\bibfield  {journal} {\bibinfo  {journal} {arXiv:1910.10745}\
  } (\bibinfo {year} {2019})}\BibitemShut {NoStop}%
\bibitem [{\citenamefont {Wilczek}(2012)}]{wilc12}%
  \BibitemOpen
  \bibfield  {author} {\bibinfo {author} {\bibfnamefont {F.}~\bibnamefont
  {Wilczek}},\ }\href {\doibase 10.1103/PhysRevLett.109.160401} {\bibfield
  {journal} {\bibinfo  {journal} {Phys. Rev. Lett.}\ }\textbf {\bibinfo
  {volume} {109}},\ \bibinfo {pages} {160401} (\bibinfo {year}
  {2012})}\BibitemShut {NoStop}%
\bibitem [{\citenamefont {Bruno}(2013)}]{brun13}%
  \BibitemOpen
  \bibfield  {author} {\bibinfo {author} {\bibfnamefont {P.}~\bibnamefont
  {Bruno}},\ }\href {\doibase 10.1103/PhysRevLett.111.070402} {\bibfield
  {journal} {\bibinfo  {journal} {Phys. Rev. Lett.}\ }\textbf {\bibinfo
  {volume} {111}},\ \bibinfo {pages} {070402} (\bibinfo {year}
  {2013})}\BibitemShut {NoStop}%
\bibitem [{\citenamefont {Watanabe}\ and\ \citenamefont
  {Oshikawa}(2015)}]{wata15}%
  \BibitemOpen
  \bibfield  {author} {\bibinfo {author} {\bibfnamefont {H.}~\bibnamefont
  {Watanabe}}\ and\ \bibinfo {author} {\bibfnamefont {M.}~\bibnamefont
  {Oshikawa}},\ }\href {\doibase 10.1103/PhysRevLett.114.251603} {\bibfield
  {journal} {\bibinfo  {journal} {Phys. Rev. Lett.}\ }\textbf {\bibinfo
  {volume} {114}},\ \bibinfo {pages} {251603} (\bibinfo {year}
  {2015})}\BibitemShut {NoStop}%
\bibitem [{\citenamefont {Kozin}\ and\ \citenamefont
  {Kyriienko}(2019)}]{kozi19}%
  \BibitemOpen
  \bibfield  {author} {\bibinfo {author} {\bibfnamefont {V.~K.}\ \bibnamefont
  {Kozin}}\ and\ \bibinfo {author} {\bibfnamefont {O.}~\bibnamefont
  {Kyriienko}},\ }\href {\doibase 10.1103/PhysRevLett.123.210602} {\bibfield
  {journal} {\bibinfo  {journal} {Phys. Rev. Lett.}\ }\textbf {\bibinfo
  {volume} {123}},\ \bibinfo {pages} {210602} (\bibinfo {year}
  {2019})}\BibitemShut {NoStop}%
\bibitem [{\citenamefont {Sacha}(2015)}]{sach15}%
  \BibitemOpen
  \bibfield  {author} {\bibinfo {author} {\bibfnamefont {K.}~\bibnamefont
  {Sacha}},\ }\href {\doibase 10.1103/PhysRevA.91.033617} {\bibfield  {journal}
  {\bibinfo  {journal} {Phys. Rev. A}\ }\textbf {\bibinfo {volume} {91}},\
  \bibinfo {pages} {033617} (\bibinfo {year} {2015})}\BibitemShut {NoStop}%
\bibitem [{\citenamefont {Khemani}\ \emph {et~al.}(2016)\citenamefont
  {Khemani}, \citenamefont {Lazarides}, \citenamefont {Moessner},\ and\
  \citenamefont {Sondhi}}]{khem16}%
  \BibitemOpen
  \bibfield  {author} {\bibinfo {author} {\bibfnamefont {V.}~\bibnamefont
  {Khemani}}, \bibinfo {author} {\bibfnamefont {A.}~\bibnamefont {Lazarides}},
  \bibinfo {author} {\bibfnamefont {R.}~\bibnamefont {Moessner}}, \ and\
  \bibinfo {author} {\bibfnamefont {S.~L.}\ \bibnamefont {Sondhi}},\ }\href
  {\doibase 10.1103/PhysRevLett.116.250401} {\bibfield  {journal} {\bibinfo
  {journal} {Phys. Rev. Lett.}\ }\textbf {\bibinfo {volume} {116}},\ \bibinfo
  {pages} {250401} (\bibinfo {year} {2016})}\BibitemShut {NoStop}%
\bibitem [{\citenamefont {Else}\ \emph {et~al.}(2016)\citenamefont {Else},
  \citenamefont {Bauer},\ and\ \citenamefont {Nayak}}]{else16}%
  \BibitemOpen
  \bibfield  {author} {\bibinfo {author} {\bibfnamefont {D.~V.}\ \bibnamefont
  {Else}}, \bibinfo {author} {\bibfnamefont {B.}~\bibnamefont {Bauer}}, \ and\
  \bibinfo {author} {\bibfnamefont {C.}~\bibnamefont {Nayak}},\ }\href
  {\doibase 10.1103/PhysRevLett.117.090402} {\bibfield  {journal} {\bibinfo
  {journal} {Phys. Rev. Lett.}\ }\textbf {\bibinfo {volume} {117}},\ \bibinfo
  {pages} {090402} (\bibinfo {year} {2016})}\BibitemShut {NoStop}%
\bibitem [{\citenamefont {Pizzi}\ \emph {et~al.}(2019)\citenamefont {Pizzi},
  \citenamefont {Knolle},\ and\ \citenamefont {Nunnenkamp}}]{pizz19}%
  \BibitemOpen
  \bibfield  {author} {\bibinfo {author} {\bibfnamefont {A.}~\bibnamefont
  {Pizzi}}, \bibinfo {author} {\bibfnamefont {J.}~\bibnamefont {Knolle}}, \
  and\ \bibinfo {author} {\bibfnamefont {A.}~\bibnamefont {Nunnenkamp}},\
  }\href {\doibase 10.1103/PhysRevLett.123.150601} {\bibfield  {journal}
  {\bibinfo  {journal} {Phys. Rev. Lett.}\ }\textbf {\bibinfo {volume} {123}},\
  \bibinfo {pages} {150601} (\bibinfo {year} {2019})}\BibitemShut {NoStop}%
\bibitem [{\citenamefont {Moessner}\ and\ \citenamefont
  {Sondhi}(2017)}]{moes17}%
  \BibitemOpen
  \bibfield  {author} {\bibinfo {author} {\bibfnamefont {R.}~\bibnamefont
  {Moessner}}\ and\ \bibinfo {author} {\bibfnamefont {S.}~\bibnamefont
  {Sondhi}},\ }\href@noop {} {\bibfield  {journal} {\bibinfo  {journal} {Nature
  Phys.}\ }\textbf {\bibinfo {volume} {13}},\ \bibinfo {pages} {424} (\bibinfo
  {year} {2017})}\BibitemShut {NoStop}%
\bibitem [{\citenamefont {Zhang}\ \emph {et~al.}(2017)\citenamefont {Zhang},
  \citenamefont {Hess}, \citenamefont {Kyprianidis}, \citenamefont {Becker},
  \citenamefont {Lee}, \citenamefont {Smith}, \citenamefont {Pagano},
  \citenamefont {Potirniche}, \citenamefont {Potter}, \citenamefont
  {Vishwanath} \emph {et~al.}}]{zhan17}%
  \BibitemOpen
  \bibfield  {author} {\bibinfo {author} {\bibfnamefont {J.}~\bibnamefont
  {Zhang}}, \bibinfo {author} {\bibfnamefont {P.}~\bibnamefont {Hess}},
  \bibinfo {author} {\bibfnamefont {A.}~\bibnamefont {Kyprianidis}}, \bibinfo
  {author} {\bibfnamefont {P.}~\bibnamefont {Becker}}, \bibinfo {author}
  {\bibfnamefont {A.}~\bibnamefont {Lee}}, \bibinfo {author} {\bibfnamefont
  {J.}~\bibnamefont {Smith}}, \bibinfo {author} {\bibfnamefont
  {G.}~\bibnamefont {Pagano}}, \bibinfo {author} {\bibfnamefont {I.-D.}\
  \bibnamefont {Potirniche}}, \bibinfo {author} {\bibfnamefont {A.~C.}\
  \bibnamefont {Potter}}, \bibinfo {author} {\bibfnamefont {A.}~\bibnamefont
  {Vishwanath}},  \emph {et~al.},\ }\href@noop {} {\bibfield  {journal}
  {\bibinfo  {journal} {Nature}\ }\textbf {\bibinfo {volume} {543}},\ \bibinfo
  {pages} {217} (\bibinfo {year} {2017})}\BibitemShut {NoStop}%
\bibitem [{\citenamefont {Choi}\ \emph {et~al.}(2017)\citenamefont {Choi},
  \citenamefont {Choi}, \citenamefont {Landig}, \citenamefont {Kucsko},
  \citenamefont {Zhou}, \citenamefont {Isoya}, \citenamefont {Jelezko},
  \citenamefont {Onoda}, \citenamefont {Sumiya}, \citenamefont {Khemani} \emph
  {et~al.}}]{choi17}%
  \BibitemOpen
  \bibfield  {author} {\bibinfo {author} {\bibfnamefont {S.}~\bibnamefont
  {Choi}}, \bibinfo {author} {\bibfnamefont {J.}~\bibnamefont {Choi}}, \bibinfo
  {author} {\bibfnamefont {R.}~\bibnamefont {Landig}}, \bibinfo {author}
  {\bibfnamefont {G.}~\bibnamefont {Kucsko}}, \bibinfo {author} {\bibfnamefont
  {H.}~\bibnamefont {Zhou}}, \bibinfo {author} {\bibfnamefont {J.}~\bibnamefont
  {Isoya}}, \bibinfo {author} {\bibfnamefont {F.}~\bibnamefont {Jelezko}},
  \bibinfo {author} {\bibfnamefont {S.}~\bibnamefont {Onoda}}, \bibinfo
  {author} {\bibfnamefont {H.}~\bibnamefont {Sumiya}}, \bibinfo {author}
  {\bibfnamefont {V.}~\bibnamefont {Khemani}},  \emph {et~al.},\ }\href@noop {}
  {\bibfield  {journal} {\bibinfo  {journal} {Nature}\ }\textbf {\bibinfo
  {volume} {543}},\ \bibinfo {pages} {221} (\bibinfo {year}
  {2017})}\BibitemShut {NoStop}%
\bibitem [{\citenamefont {Lazarides}\ and\ \citenamefont
  {Moessner}(2017)}]{laza17}%
  \BibitemOpen
  \bibfield  {author} {\bibinfo {author} {\bibfnamefont {A.}~\bibnamefont
  {Lazarides}}\ and\ \bibinfo {author} {\bibfnamefont {R.}~\bibnamefont
  {Moessner}},\ }\href {\doibase 10.1103/PhysRevB.95.195135} {\bibfield
  {journal} {\bibinfo  {journal} {Phys. Rev. B}\ }\textbf {\bibinfo {volume}
  {95}},\ \bibinfo {pages} {195135} (\bibinfo {year} {2017})}\BibitemShut
  {NoStop}%
\bibitem [{\citenamefont {Yao}\ \emph {et~al.}(2018)\citenamefont {Yao},
  \citenamefont {Nayak}, \citenamefont {Balents},\ and\ \citenamefont
  {Zaletel}}]{yao18}%
  \BibitemOpen
  \bibfield  {author} {\bibinfo {author} {\bibfnamefont {N.~Y.}\ \bibnamefont
  {Yao}}, \bibinfo {author} {\bibfnamefont {C.}~\bibnamefont {Nayak}}, \bibinfo
  {author} {\bibfnamefont {L.}~\bibnamefont {Balents}}, \ and\ \bibinfo
  {author} {\bibfnamefont {M.~P.}\ \bibnamefont {Zaletel}},\ }\href@noop {}
  {\bibfield  {journal} {\bibinfo  {journal} {arXiv:1801.02628}\ } (\bibinfo
  {year} {2018})}\BibitemShut {NoStop}%
\bibitem [{\citenamefont {Gong}\ \emph {et~al.}(2018)\citenamefont {Gong},
  \citenamefont {Hamazaki},\ and\ \citenamefont {Ueda}}]{gong18}%
  \BibitemOpen
  \bibfield  {author} {\bibinfo {author} {\bibfnamefont {Z.}~\bibnamefont
  {Gong}}, \bibinfo {author} {\bibfnamefont {R.}~\bibnamefont {Hamazaki}}, \
  and\ \bibinfo {author} {\bibfnamefont {M.}~\bibnamefont {Ueda}},\ }\href
  {\doibase 10.1103/PhysRevLett.120.040404} {\bibfield  {journal} {\bibinfo
  {journal} {Phys. Rev. Lett.}\ }\textbf {\bibinfo {volume} {120}},\ \bibinfo
  {pages} {040404} (\bibinfo {year} {2018})}\BibitemShut {NoStop}%
\bibitem [{\citenamefont {Wang}\ \emph {et~al.}(2018)\citenamefont {Wang},
  \citenamefont {Xing}, \citenamefont {Carlo},\ and\ \citenamefont
  {Poletti}}]{wang18}%
  \BibitemOpen
  \bibfield  {author} {\bibinfo {author} {\bibfnamefont {R.~R.~W.}\
  \bibnamefont {Wang}}, \bibinfo {author} {\bibfnamefont {B.}~\bibnamefont
  {Xing}}, \bibinfo {author} {\bibfnamefont {G.~G.}\ \bibnamefont {Carlo}}, \
  and\ \bibinfo {author} {\bibfnamefont {D.}~\bibnamefont {Poletti}},\ }\href
  {\doibase 10.1103/PhysRevE.97.020202} {\bibfield  {journal} {\bibinfo
  {journal} {Phys. Rev. E}\ }\textbf {\bibinfo {volume} {97}},\ \bibinfo
  {pages} {020202} (\bibinfo {year} {2018})}\BibitemShut {NoStop}%
\bibitem [{\citenamefont {Gambetta}\ \emph
  {et~al.}(2019{\natexlab{a}})\citenamefont {Gambetta}, \citenamefont
  {Carollo}, \citenamefont {Marcuzzi}, \citenamefont {Garrahan},\ and\
  \citenamefont {Lesanovsky}}]{gamb19}%
  \BibitemOpen
  \bibfield  {author} {\bibinfo {author} {\bibfnamefont {F.~M.}\ \bibnamefont
  {Gambetta}}, \bibinfo {author} {\bibfnamefont {F.}~\bibnamefont {Carollo}},
  \bibinfo {author} {\bibfnamefont {M.}~\bibnamefont {Marcuzzi}}, \bibinfo
  {author} {\bibfnamefont {J.~P.}\ \bibnamefont {Garrahan}}, \ and\ \bibinfo
  {author} {\bibfnamefont {I.}~\bibnamefont {Lesanovsky}},\ }\href {\doibase
  10.1103/PhysRevLett.122.015701} {\bibfield  {journal} {\bibinfo  {journal}
  {Phys. Rev. Lett.}\ }\textbf {\bibinfo {volume} {122}},\ \bibinfo {pages}
  {015701} (\bibinfo {year} {2019}{\natexlab{a}})}\BibitemShut {NoStop}%
\bibitem [{\citenamefont {Gambetta}\ \emph
  {et~al.}(2019{\natexlab{b}})\citenamefont {Gambetta}, \citenamefont
  {Carollo}, \citenamefont {Lazarides}, \citenamefont {Lesanovsky},\ and\
  \citenamefont {Garrahan}}]{gamb19b}%
  \BibitemOpen
  \bibfield  {author} {\bibinfo {author} {\bibfnamefont {F.~M.}\ \bibnamefont
  {Gambetta}}, \bibinfo {author} {\bibfnamefont {F.}~\bibnamefont {Carollo}},
  \bibinfo {author} {\bibfnamefont {A.}~\bibnamefont {Lazarides}}, \bibinfo
  {author} {\bibfnamefont {I.}~\bibnamefont {Lesanovsky}}, \ and\ \bibinfo
  {author} {\bibfnamefont {J.~P.}\ \bibnamefont {Garrahan}},\ }\href {\doibase
  10.1103/PhysRevE.100.060105} {\bibfield  {journal} {\bibinfo  {journal}
  {Phys. Rev. E}\ }\textbf {\bibinfo {volume} {100}},\ \bibinfo {pages}
  {060105} (\bibinfo {year} {2019}{\natexlab{b}})}\BibitemShut {NoStop}%
\bibitem [{\citenamefont {Heugel}\ \emph {et~al.}(2019)\citenamefont {Heugel},
  \citenamefont {Oscity}, \citenamefont {Eichler}, \citenamefont {Zilberberg},\
  and\ \citenamefont {Chitra}}]{heug19}%
  \BibitemOpen
  \bibfield  {author} {\bibinfo {author} {\bibfnamefont {T.~L.}\ \bibnamefont
  {Heugel}}, \bibinfo {author} {\bibfnamefont {M.}~\bibnamefont {Oscity}},
  \bibinfo {author} {\bibfnamefont {A.}~\bibnamefont {Eichler}}, \bibinfo
  {author} {\bibfnamefont {O.}~\bibnamefont {Zilberberg}}, \ and\ \bibinfo
  {author} {\bibfnamefont {R.}~\bibnamefont {Chitra}},\ }\href {\doibase
  10.1103/PhysRevLett.123.124301} {\bibfield  {journal} {\bibinfo  {journal}
  {Phys. Rev. Lett.}\ }\textbf {\bibinfo {volume} {123}},\ \bibinfo {pages}
  {124301} (\bibinfo {year} {2019})}\BibitemShut {NoStop}%
\bibitem [{\citenamefont {Goldstein}(2018)}]{gold18}%
  \BibitemOpen
  \bibfield  {author} {\bibinfo {author} {\bibfnamefont {R.~E.}\ \bibnamefont
  {Goldstein}},\ }\href@noop {} {\bibfield  {journal} {\bibinfo  {journal}
  {Physics Today}\ }\textbf {\bibinfo {volume} {71}},\ \bibinfo {pages} {32}
  (\bibinfo {year} {2018})}\BibitemShut {NoStop}%
\bibitem [{\citenamefont {Seifert}(2012)}]{seif12}%
  \BibitemOpen
  \bibfield  {author} {\bibinfo {author} {\bibfnamefont {U.}~\bibnamefont
  {Seifert}},\ }\href {\doibase 10.1088/0034-4885/75/12/126001} {\bibfield
  {journal} {\bibinfo  {journal} {Rep. Prog. Phys.}\ }\textbf {\bibinfo
  {volume} {75}},\ \bibinfo {pages} {126001} (\bibinfo {year}
  {2012})}\BibitemShut {NoStop}%
\bibitem [{\citenamefont {{J}arzynski}(2011)}]{jarz11}%
  \BibitemOpen
  \bibfield  {author} {\bibinfo {author} {\bibfnamefont {C.}~\bibnamefont
  {{J}arzynski}},\ }\href {\doibase 10.1146/annurev-conmatphys-062910-140506}
  {\bibfield  {journal} {\bibinfo  {journal} {Ann. Rev. Cond. Mat. Phys.}\
  }\textbf {\bibinfo {volume} {2}},\ \bibinfo {pages} {329} (\bibinfo {year}
  {2011})}\BibitemShut {NoStop}%
\bibitem [{\citenamefont {den Broeck}\ and\ \citenamefont
  {Esposito}(2015)}]{broe15}%
  \BibitemOpen
  \bibfield  {author} {\bibinfo {author} {\bibfnamefont {C.~V.}\ \bibnamefont
  {den Broeck}}\ and\ \bibinfo {author} {\bibfnamefont {M.}~\bibnamefont
  {Esposito}},\ }\href {\doibase https://doi.org/10.1016/j.physa.2014.04.035}
  {\bibfield  {journal} {\bibinfo  {journal} {Physica A}\ }\textbf {\bibinfo
  {volume} {418}},\ \bibinfo {pages} {6} (\bibinfo {year} {2015})}\BibitemShut
  {NoStop}%
\bibitem [{\citenamefont {Oberreiter}\ \emph {et~al.}(2019)\citenamefont
  {Oberreiter}, \citenamefont {Seifert},\ and\ \citenamefont
  {Barato}}]{ober19}%
  \BibitemOpen
  \bibfield  {author} {\bibinfo {author} {\bibfnamefont {L.}~\bibnamefont
  {Oberreiter}}, \bibinfo {author} {\bibfnamefont {U.}~\bibnamefont {Seifert}},
  \ and\ \bibinfo {author} {\bibfnamefont {A.~C.}\ \bibnamefont {Barato}},\
  }\href {\doibase 10.1103/PhysRevE.100.012135} {\bibfield  {journal} {\bibinfo
   {journal} {Phys. Rev. E}\ }\textbf {\bibinfo {volume} {100}},\ \bibinfo
  {pages} {012135} (\bibinfo {year} {2019})}\BibitemShut {NoStop}%
\bibitem [{\citenamefont {Gupta}\ \emph {et~al.}(2018)\citenamefont {Gupta},
  \citenamefont {Campa},\ and\ \citenamefont {Ruffo}}]{gupt18}%
  \BibitemOpen
  \bibfield  {author} {\bibinfo {author} {\bibfnamefont {S.}~\bibnamefont
  {Gupta}}, \bibinfo {author} {\bibfnamefont {A.}~\bibnamefont {Campa}}, \ and\
  \bibinfo {author} {\bibfnamefont {S.}~\bibnamefont {Ruffo}},\ }\href@noop {}
  {\emph {\bibinfo {title} {Statistical Physics of Synchronization}}}\
  (\bibinfo  {publisher} {Springer},\ \bibinfo {year} {2018})\BibitemShut
  {NoStop}%
\bibitem [{\citenamefont {Wood}\ \emph
  {et~al.}(2006{\natexlab{a}})\citenamefont {Wood}, \citenamefont {Van~den
  Broeck}, \citenamefont {Kawai},\ and\ \citenamefont {Lindenberg}}]{wood06}%
  \BibitemOpen
  \bibfield  {author} {\bibinfo {author} {\bibfnamefont {K.}~\bibnamefont
  {Wood}}, \bibinfo {author} {\bibfnamefont {C.}~\bibnamefont {Van~den
  Broeck}}, \bibinfo {author} {\bibfnamefont {R.}~\bibnamefont {Kawai}}, \ and\
  \bibinfo {author} {\bibfnamefont {K.}~\bibnamefont {Lindenberg}},\ }\href
  {\doibase 10.1103/PhysRevLett.96.145701} {\bibfield  {journal} {\bibinfo
  {journal} {Phys. Rev. Lett.}\ }\textbf {\bibinfo {volume} {96}},\ \bibinfo
  {pages} {145701} (\bibinfo {year} {2006}{\natexlab{a}})}\BibitemShut
  {NoStop}%
\bibitem [{\citenamefont {Wood}\ \emph
  {et~al.}(2006{\natexlab{b}})\citenamefont {Wood}, \citenamefont {Van~den
  Broeck}, \citenamefont {Kawai},\ and\ \citenamefont {Lindenberg}}]{wood06a}%
  \BibitemOpen
  \bibfield  {author} {\bibinfo {author} {\bibfnamefont {K.}~\bibnamefont
  {Wood}}, \bibinfo {author} {\bibfnamefont {C.}~\bibnamefont {Van~den
  Broeck}}, \bibinfo {author} {\bibfnamefont {R.}~\bibnamefont {Kawai}}, \ and\
  \bibinfo {author} {\bibfnamefont {K.}~\bibnamefont {Lindenberg}},\ }\href
  {\doibase 10.1103/PhysRevE.74.031113} {\bibfield  {journal} {\bibinfo
  {journal} {Phys. Rev. E}\ }\textbf {\bibinfo {volume} {74}},\ \bibinfo
  {pages} {031113} (\bibinfo {year} {2006}{\natexlab{b}})}\BibitemShut
  {NoStop}%
\bibitem [{\citenamefont {Gillespie}(1977)}]{gill77}%
  \BibitemOpen
  \bibfield  {author} {\bibinfo {author} {\bibfnamefont {D.~T.}\ \bibnamefont
  {Gillespie}},\ }\href {\doibase 10.1021/j100540a008} {\bibfield  {journal}
  {\bibinfo  {journal} {J.\ Phys.\ Chem.}\ }\textbf {\bibinfo {volume} {81}},\
  \bibinfo {pages} {2340} (\bibinfo {year} {1977})}\BibitemShut {NoStop}%
\bibitem [{\citenamefont {Herpich}\ \emph {et~al.}(2018)\citenamefont
  {Herpich}, \citenamefont {Thingna},\ and\ \citenamefont {Esposito}}]{herp18}%
  \BibitemOpen
  \bibfield  {author} {\bibinfo {author} {\bibfnamefont {T.}~\bibnamefont
  {Herpich}}, \bibinfo {author} {\bibfnamefont {J.}~\bibnamefont {Thingna}}, \
  and\ \bibinfo {author} {\bibfnamefont {M.}~\bibnamefont {Esposito}},\ }\href
  {\doibase 10.1103/PhysRevX.8.031056} {\bibfield  {journal} {\bibinfo
  {journal} {Phys. Rev. X}\ }\textbf {\bibinfo {volume} {8}},\ \bibinfo {pages}
  {031056} (\bibinfo {year} {2018})}\BibitemShut {NoStop}%
\bibitem [{\citenamefont {Nguyen}\ \emph {et~al.}(2018)\citenamefont {Nguyen},
  \citenamefont {Seifert},\ and\ \citenamefont {Barato}}]{nguy18}%
  \BibitemOpen
  \bibfield  {author} {\bibinfo {author} {\bibfnamefont {B.}~\bibnamefont
  {Nguyen}}, \bibinfo {author} {\bibfnamefont {U.}~\bibnamefont {Seifert}}, \
  and\ \bibinfo {author} {\bibfnamefont {A.~C.}\ \bibnamefont {Barato}},\
  }\href {\doibase 10.1063/1.5032104} {\bibfield  {journal} {\bibinfo
  {journal} {J.\ Chem.\ Phys.}\ }\textbf {\bibinfo {volume} {149}},\ \bibinfo
  {pages} {045101} (\bibinfo {year} {2018})}\BibitemShut {NoStop}%
\bibitem [{\citenamefont {Esposito}\ \emph {et~al.}(2009)\citenamefont
  {Esposito}, \citenamefont {Lindenberg},\ and\ \citenamefont {van~den
  Broeck}}]{espo09}%
  \BibitemOpen
  \bibfield  {author} {\bibinfo {author} {\bibfnamefont {M.}~\bibnamefont
  {Esposito}}, \bibinfo {author} {\bibfnamefont {K.}~\bibnamefont
  {Lindenberg}}, \ and\ \bibinfo {author} {\bibfnamefont {C.}~\bibnamefont
  {van~den Broeck}},\ }\href {\doibase 10.1103/PhysRevLett.102.130602}
  {\bibfield  {journal} {\bibinfo  {journal} {Phys. Rev. Lett.}\ }\textbf
  {\bibinfo {volume} {102}},\ \bibinfo {pages} {130602} (\bibinfo {year}
  {2009})}\BibitemShut {NoStop}%
\bibitem [{\citenamefont {Shiraishi}\ \emph {et~al.}(2016)\citenamefont
  {Shiraishi}, \citenamefont {Saito},\ and\ \citenamefont {Tasaki}}]{shir16}%
  \BibitemOpen
  \bibfield  {author} {\bibinfo {author} {\bibfnamefont {N.}~\bibnamefont
  {Shiraishi}}, \bibinfo {author} {\bibfnamefont {K.}~\bibnamefont {Saito}}, \
  and\ \bibinfo {author} {\bibfnamefont {H.}~\bibnamefont {Tasaki}},\ }\href
  {\doibase 10.1103/PhysRevLett.117.190601} {\bibfield  {journal} {\bibinfo
  {journal} {Phys. Rev. Lett.}\ }\textbf {\bibinfo {volume} {117}},\ \bibinfo
  {pages} {190601} (\bibinfo {year} {2016})}\BibitemShut {NoStop}%
\bibitem [{\citenamefont {Pietzonka}\ and\ \citenamefont
  {Seifert}(2018)}]{piet18}%
  \BibitemOpen
  \bibfield  {author} {\bibinfo {author} {\bibfnamefont {P.}~\bibnamefont
  {Pietzonka}}\ and\ \bibinfo {author} {\bibfnamefont {U.}~\bibnamefont
  {Seifert}},\ }\href@noop {} {\bibfield  {journal} {\bibinfo  {journal} {Phys.
  Rev. Lett.}\ }\textbf {\bibinfo {volume} {120}},\ \bibinfo {pages} {190602}
  (\bibinfo {year} {2018})}\BibitemShut {NoStop}%
\bibitem{koyu19} T. Koyuk and U. Seifert, Phys. Rev. Lett. \textbf{122}, 230601 (2019).
\end{thebibliography}%

\end{document}